\documentclass[aps,amsfonts,amsmath,prd,preprint,nofootinbib]{revtex4}
\pdfoutput=1
\newcommand{\beq}{\begin{equation}}
\newcommand{\eeq}{\end{equation}}
\usepackage{graphicx}
\usepackage{natbib}
\usepackage[utf8]{inputenc}
\begin{document}

\title{Primordial black hole formation by vacuum bubbles II}
\author{Heling Deng}
\email{dengheling@gmail.com}
\affiliation{Physics Department, Arizona State University, Tempe, AZ 85287, USA}

\begin{abstract}

The discoveries of LIGO/Virgo black holes in recent years have revitalized the study of primordial black holes. In this work we investigate a mechanism where primordial black holes are formed by vacuum bubbles that randomly nucleate during inflation through quantum tunneling. After inflation, these bubbles typically run into the ambient radiation fluid with a large Lorentz factor. In our previous work, we assumed the bubble fields are strongly coupled to the standard model particles so that the bubble wall is impermeable. Here we complete this picture by considering bubbles interacting with the fluid only through gravity. By studying the scenario in several limits, we found that black holes could form in the either subcritical or supercritical regime. Depending on the model parameters, the resulting mass spectrum of the black holes could be wide or narrow, and may develop two peaks separated by a large mass range. With different spectra, these black holes may account for the LIGO/Virgo black holes, supermassive black holes, and may play an important role in dark matter.

\end{abstract}

\maketitle

\section{Introduction}

Primordial black holes (PBHs) are hypothetical black holes formed
in the early universe, usually during the radiation dominated era. In principle, a PBH could have a mass ranging from the
Planck mass ($M_{\rm Pl}\sim10^{-5}\ \text{g}$) to many orders of magnitude larger than the
solar mass $(M_{\odot}\sim10^{33}\ \text{g})$.  Due to Hawking radiation, PBHs formed with masses below $\sim 10^{15}\ \text{g}$
would have evaporated by now, hence many efforts to constrain them have been focused on nonevaporating ones with $M \gtrsim 10^{15}\ \text{g}$.

Recent interest in PBHs is to a great extent stimulated by the discoveries
in LIGO/Virgo: gravitational waves were detected from merging black holes of masses $\mathcal{O}(10\mbox{--}100)M_{\odot}$ \cite{LIGOScientific:2018mvr}. These black holes are slightly heavier than what would be expected for ordinary stellar black holes, and one intriguing hypothesis is that they have a primordial origin \cite{Bird:2016dcv,Clesse:2016vqa,Sasaki:2016jop}. 

\begin{figure}
\includegraphics[scale=0.3]{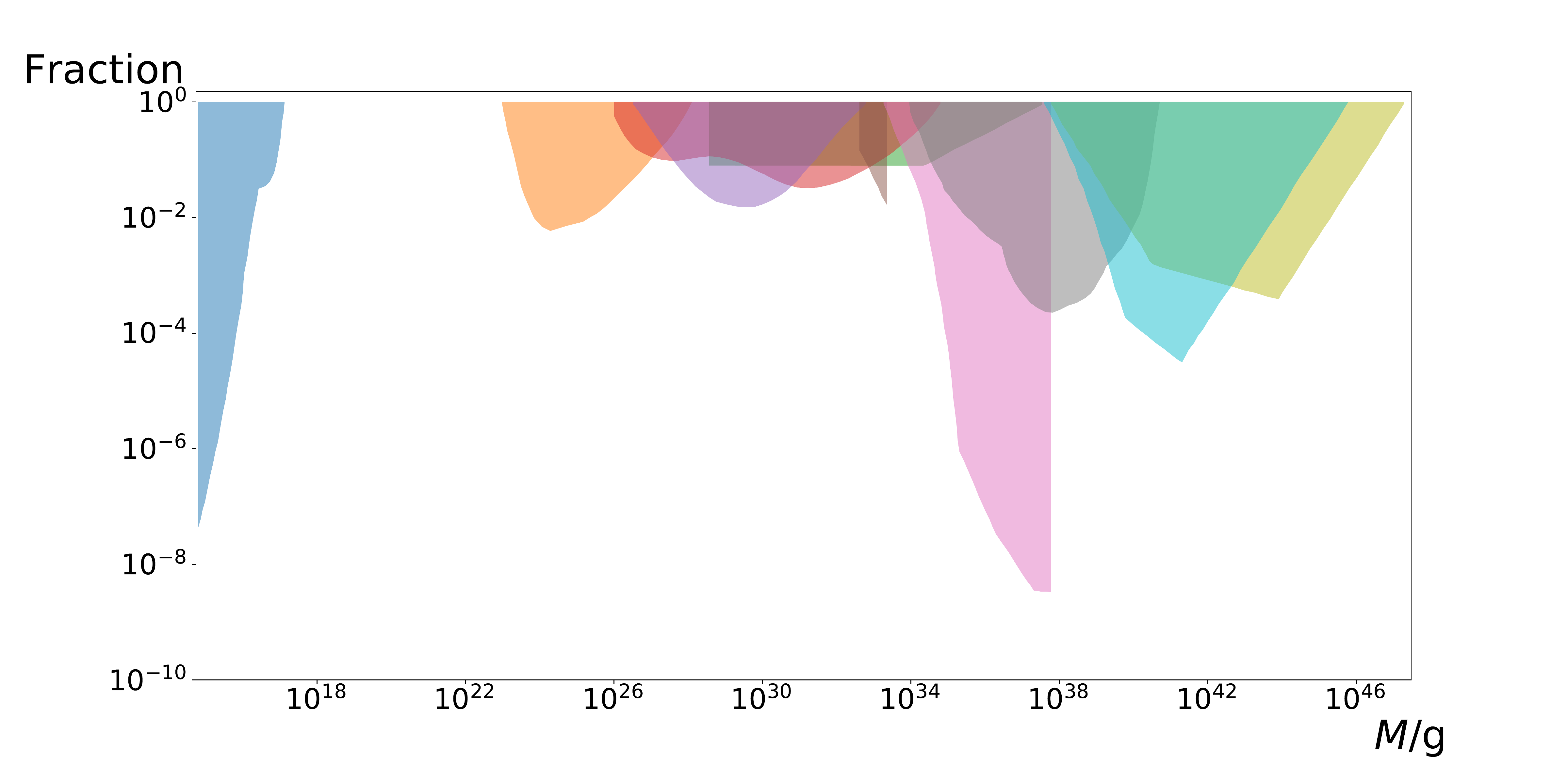}

\caption{\label{fig:f_max}Some conservative constraints on the fraction of
dark matter in PBHs for a monochromatic mass spectrum (see ref. \cite{Carr:2020gox} and references therein). Colored regions have been excluded by observations including
(from left to right) evaporation (blue), HSC (orange), EROS (red),
OGLE (purple), Icarus (green), LIGO/Virgo (brown), Planck (pink),
X-ray binaries (gray), halo dynamical friction (cyan), and large
scale structure (olive). }

\end{figure}

LIGO/Virgo events also revitalized the study on PBH dark matter (see ref. \cite{Carr:2020xqk} for a recent review). At the moment, there are stringent constraints on the fraction of dark matter in PBHs within the mass range $10^{15}\mbox{--}10^{47}\ \text{g}$ from the observed dynamical, microlensing and astrophysical effects (see refs. \cite{Sasaki:2018dmp, Carr:2020gox} and references therein). Some up-to-date upper
bounds $f_{max}(M)$ for a monochromatic mass spectrum are shown in fig. \ref{fig:f_max} (adapted from fig. 10 in ref. \cite{Carr:2020gox}), with colored regions ruled out by observations. The possibility that PBHs account for all dark matter has been excluded for most black hole masses. In fact, LIGO/Virgo is among the observations to recently close one of the windows: if PBHs play any role in dark matter, they can only constitute a small fraction at around $10M_{\odot}$, otherwise
the black hole merger rate inferred by LIGO/Virgo should have been
larger \cite{Vaskonen:2019jpv}. Another tight bound imposed recently is from the study of
the CMB anisotropy in ref. \cite{Serpico:2020ehh}, which considers the effect of disk and spherical accretion
of a halo around PBHs, and strongly constrains the population of
PBHs in the mass range around $1\mbox{--}10^{4}\ M_{\odot}$, reaching
$f_{max}(10^{4}M_{\odot})<3\times10^{-9}$. We can see from fig.
\ref{fig:f_max} that the only window allowing PBHs to constitute
all dark matter is $\sim10^{17}\mbox{--}10^{23}\ \text{g}$.

Another motivation to the study of PBHs is to explain supermassive
black holes (SMBHs) at the center of most galaxies \cite{LyndenBell:1969yx, Kormendy:1995er}. These
black holes have masses ranging from around $10^{6}M_{\odot}$ to
$10^{10}M_{\odot}$, which cannot be explained by standard accretion
models \cite{Haiman:2004ve}, and observations of quasars indicate that many of them were already in place at high redshifts (see ref. \cite{Kormendy:2013dxa} for a review). One is then led to speculate that SMBHs were seeded by PBHs, which could have large masses by birth \cite{Rubin:2001yw, Bean:2002kx, Duechting:2004dk, Clesse:2015wea, Carr:2018rid}. It was shown in ref. \cite{Duechting:2004dk,Serpico:2020ehh} that primordial seeds as massive as $\sim10^{3}M_{\odot}$ can attain sufficient accretion and grow to SMBHs.

PBHs can be formed in a variety of mechanisms. The most popular and natural scenario is PBH formation from inflationary density perturbations \cite{Ivanov:1994pa, GarciaBellido:1996qt, Kawasaki:1997ju, Yokoyama:1998pt, Garcia-Bellido:2017mdw, Hertzberg:2017dkh}: after inflation, a large overdensity of superhorizonal scale can overcome pressure and collapse into
a black hole at Hubble reentry. However, provided the primordial perturbations are Gaussian, the formation of these black holes with masses $\sim10^{5}\mbox{--}10^{9}\ M_{\odot}$ has been ruled out in any appreciable numbers due to the strong bounds
on $\mu$-distortions in CMB \cite{Chluba:2012gq, Chluba:2012we, Kohri:2014lza, Nakama:2017xvq}.
In order to account for the seeds of SMBHs, one thus needs an inflationary model that happens to give a relatively large density of PBHs with $M\sim10^{3}M_{\odot}$ and a negligible density at $M\sim10^{5}M_{\odot}$.
Other ways to circumvent this problem include considering perturbations with a
highly non-Gaussian tail (e.g., \cite{Nakama:2016kfq}), and other PBH mechanisms unrelated to density perturbations from quantum fluctuations (e.g., collapse of cosmic strings~\cite{Hawking:1987bn, Polnarev:1988dh, Garriga:1992nm, Caldwell:1995fu}, bubble collisions~\cite{Hawking:1982ga,Khlopov:1999ys,Dymnikova:2000dy}, and collapse of closed domain walls \cite{Garriga:1992nm,Rubin:2000dq, Khlopov:2004sc}).

In this paper we shall turn to the last approach and investigate a
simple and natural mechanism, where the black holes are formed by
vacuum bubbles that nucleate through quantum tunneling during inflation. Similar models were
constructed and analyzed in ref. \cite{Garriga:2015fdk}, and studied in detail by numerical work in refs. \cite{Deng:2016vzb,Deng:2017uwc}, where we focused on two specific scenarios: (1) permeable spherical domain walls immersed
in the radiation dominated universe, and (2) vacuum bubbles with impermeable walls
that completely reflect the radiation fluid after inflation. Black holes could be formed in either case with a distinctive and extended mass distribution. In the second scenario, we assumed strong couplings between the fields in the standard model and those producing the bubble. As such, standard model particles could acquire large masses in the bubble interior, and particles outside bounce back when they hit the walls. Bubbles running into the ambient fluid with a large Lorentz factor would then lose much of their energy due to momentum transfer. This takes place almost instantaneously compared to the Hubble time, and the bubbles do not grow much relative to the Hubble flow before turning into black holes. In this work, we will complete
this picture and consider vacuum bubbles interacting with the FRW
fluid only through gravitation. The bubbles could then grow to much larger
sizes as the fluid freely flows in. We found that this would lead to
a drastic change in the mass spectrum of the resulting black holes,
especially on the end of the small ones. Depending on the model parameters,
the spectrum could be either wide or narrow, and could be consistent
with all the current constraints, while in the meantime being able
to play the role of dark matter, LIGO/Virgo black holes and/or SMBHs.
Our main result are shown in fig. \ref{fig:limit1}.

The rest of the paper is organized as follows. The behavior of bubble
expansion after inflation will be discussed in section \ref{Bubble evolution},
followed by the discussion of black hole formation in some particular situations
in section \ref{Black hole formation}. In section \ref{Mass spectrum} we will compute the corresponding mass
spectrum of the resulting black holes, and section \ref{Observational bounds} will show how
the spectrum is constrained by current observations. Conclusions
are summarized and discussed in section \ref{discussion}. We set $c=\hbar=G=1$,
and the Planck mass $M_{Pl}=\sqrt{\hbar c/G}$ throughout the paper.

\section{\label{Bubble evolution} Bubble expansion}

Inflation is usually described by the evolution of a scalar field
called the inflaton. The inflaton slowly rolls down its potential
at a large energy scale that remains almost a constant, which drives a
nearly exponential expansion. After the universe grows by over 30 orders of magnitude, the inflaton eventually rolls down to a
valley corresponding to our current vacuum. In general, the inflaton
may run in a multi-dimensional potential, including a number of other
vacua separated by barriers. Due to the effect of quantum tunneling,
a bubble may nucleate, with the inflaton in a small, spherical region
transiting to a new vacuum, where the energy scale is smaller than
the inflationary scale, while the exterior region continues its quasi-de Sitter expansion. 

Let $\rho_{i}$ be the inflationary density, $\rho_{b}$ the interior
vacuum density and $\sigma$ the surface tension (or surface energy density) of the bubble wall,
with $\rho_{i}>\rho_{b}$. We define $H_{i}\equiv(8\pi\rho_{i}/3)^{1/2}$,
$H_{b}\equiv(8\pi\rho_{b}/3)^{1/2}$ and $H_{\sigma}\equiv2\pi\sigma$.
On dimensional grounds, $H_{i}\sim\eta_{i}^{2}/M_{Pl}$, $H_{b}\sim\eta_{b}^{2}/M_{Pl}$
and $H_{\sigma}\sim\eta_{\sigma}^{3}/M_{Pl}^{2}$, where $\eta_{i}$,
$\eta_{b}$ and $\eta_{\sigma}$ are the energy scales of the corresponding
quantities. After nucleation, the bubble expands due to pressure difference
between two sides of the wall, with a Lorentz factor approaching \cite{Berezin:1987bc}
\begin{equation}
\gamma_{i}=\frac{\left[(\rho_{i}+\rho_{b}+6\pi\sigma^{2})^{2}-4\rho_{i}\rho_{b}\right]^{1/2}}{3\sigma H_{i}}.
\end{equation}
If $\eta_{i}$ dominates over $\eta_{b}$ and $\eta_{\sigma}$, this
reduces to
\begin{equation}
\gamma_{i}\approx\frac{\rho_{i}}{3\sigma H_{i}}\sim\frac{\eta_{i}^{2}M_{Pl}}{\eta_{\sigma}^{3}}\gg1.\label{gamma}
\end{equation}
Therefore, a typical bubble would expand with a large Lorentz factor
at the end of inflation. As the inflaton rolls from the quasi-de Sitter
vacuum to ours, the shape of the barrier between the bubble interior
and the slow-roll path changes, which may cause a significant change
in the bubble wall tension. In the following we regard the wall tension
after inflation as a free parameter, and will continue calling it
$\sigma$. Hence the free parameters of the bubble are $H_{i},H_{b},H_{\sigma}$
and $\gamma_{i}$ (later we will also include the bubble nucleation
rate). 

Let $t_{i}$ be the time when inflation ends. After $t_{i}$, the
exterior vacuum thermalizes into hot radiation, and the bubble wall
runs into the ambient radiation fluid with a large $\gamma_{i}$.
We assume the thermalization occurs instantaneously so that the radiation
density at $t_{i}$ is $\rho_{i}=3/32\pi t_{i}^{2}$. If there is
no interaction between the bubble and the fluid other than gravity,
the fluid freely penetrates the wall, and the exterior spacetime can
be described by a flat FRW metric,
\begin{equation}
ds^{2}=dt^{2}-a^{2}(t)(dr^{2}+r^{2}d\Omega^{2}),
\end{equation}
where $a(t)\equiv\left(t/t_{i}\right)^{1/2}$. Since both the interior
vacuum pressure and the wall tension point inwards, the bubble would
slow down and come to a halt with respect to the Hubble flow at some
time denoted by $t_{s}$. If $\gamma_{i}$ is large and $\rho_{b}$
and $\sigma$ small, the bubble would grow significantly. To be more
specific, let $a_{s}\equiv a(t_{s})=\left(t_{s}/t_{i}\right)^{1/2}$,
we are mainly interested in the cases where $a_{s}\gg1$. If $a_{s}\sim1$,
the bubble does not grow much relative to the Hubble flow before turning around, and negligible
amount of radiation flows into the bubble. As the bubble shrinks,
an empty layer would be formed between the bubble wall and the radiation
fluid. This would then be similar to the scenario discussed in
ref. \cite{Deng:2017uwc}, where the fluid is assumed to be completely reflected by
the bubble wall and the bubble stops expanding with the fluid almost immediately
after $t_{i}$. We will see that assuming $a_{s}\gg1$ would lead
to different features in the mass spectrum of the resulting black
holes. 

Let $\chi(t)$ be the comoving radius of the bubble wall for an exterior
FRW observer. The bubble runs into the radiation at a speed close to the
speed of light, so the trajectory of the wall can be approximated
by $a\dot{\chi}\approx1$ (at least at early times), where the overdot
represents the first derivative with respect to $t$. This gives
\begin{equation}
\chi(t)\approx\chi_{i}+2\sqrt{t_{i}}(\sqrt{t}-\sqrt{t_{i}}),\label{chi}
\end{equation}
where $\chi_{i}\equiv\chi(t_{i})$. Hence the comoving radius of the
bubble wall at $t_{s}$ is approximately
\begin{equation}
\chi_{s}\equiv\chi(t_{s})\sim\chi_{i}+a_{s}H_{i}^{-1},\label{chi_s}
\end{equation}
and the physical radius is
\begin{equation}
R_{s}\equiv a(t_{s})\chi_{s}\sim a_{s}\chi_{i}+a_{s}^2 H_{i}^{-1}.\label{R_s}
\end{equation}

The equation of motion of the wall can be more precisely determined
by matching the spacetimes inside and outside the bubble (see appendix),
\begin{equation}
\ddot{\chi}+\left(4-3a^{2}\dot{\chi}^{2}\right)H\dot{\chi}+\frac{2}{a^{2}\chi}\left(1-a^{2}\dot{\chi}^{2}\right)=-\left(\frac{\rho_{b}}{\sigma}+6\pi\sigma\right)\frac{\left(1-a^{2}\dot{\chi}^{2}\right)^{3/2}}{a},\label{chiEOM}
\end{equation}
where $H\equiv\dot{a}/a=(2t)^{-1}$ is the Hubble parameter in the exterior region. In principle,
this equation is valid only before the peculiar expansion speed of
the wall $a\dot{\chi}$ decreases to the speed of sound (corresponding
to a Lorentz factor $\gamma=\sqrt{3/2}$), because after this time
information inside the bubble could affect the exterior spacetime. However, if the wall tension $\sigma$ is sufficiently small, as
assumed through out the paper, fluid nearby is barely perturbed
by the wall's repulsion when the bubble expands. In addition, we shall be interested in the
following two limits: small $\rho_{b}$ and large $\rho_{b}$. If
the interior vacuum density $\rho_{b}$ is much smaller than the radiation
density during bubble expansion, radiation inside the bubble barely
gets diluted by the vacuum energy, so that the wall basically lives
in a homogeneous background when expanding. If $\rho_{b}$ is sufficiently
large, i.e., the vacuum pressure dragging the bubble is sufficiently
large, the bubble would first expand almost at the speed of light
for a while, then turns around and acquires a large speed within a
very short time. In either limit, we expect eq. (\ref{chiEOM}) to
give a good approximation of how the Lorentz factor of the wall decreases
from $\gamma_{i}$ to $1$. 

For a bubble far beyond the Hubble horizon at $t_{i}$, eq. (\ref{chiEOM})
has an analytic solution. It is shown in the appendix that, if $\chi_{i}$
is sufficiently large ($\chi_{i}\gg H_{i}^{-1}$), the third term
on the left hand side can be neglected, and $a_{s}$ becomes a constant
independent of $\chi_{i}$ (eq. (\ref{a_s})), 
\begin{equation}
a_{s}=\left[\frac{5\gamma_{i}}{2t_{i}}\left(\frac{\rho_{b}}{\sigma}+6\pi\sigma\right)^{-1}+1\right]^{1/5}.
\end{equation}
In the case that the bubble wall tension $\sigma$ does not change significantly
after inflation, this, along with eq. (\ref{gamma}), gives
\begin{equation}
a_{s}\sim\left(\frac{H_{i}^{2}}{H_{b}^{2}+H_{\sigma}^{2}}\right)^{1/5},\label{eq:a_s_limit2}
\end{equation}
which is much larger than $1$ if $H_{i}\gg H_{b},H_{\sigma}.$ Since
the third term on the left hand side of eq. (\ref{chiEOM}) tends
to slow down the bubble wall, the actual $a_{s}$ could be significantly
smaller. For more general cases, we need to numerically solve eq.
(\ref{chiEOM}) in order to find out the bubble wall's trajectory
before $t_{s}$. With the bubble size at $t_{s}$, we will be able to find an estimate for the mass of the black hole formed by the
bubble.

\section{\label{Black hole formation} Black hole formation}

After $t_{s}$, the bubble turns around for an exterior FRW observer and interacts
with the interior inhomogeneous fluid, which has been influenced by
the bubble wall and the interior vacuum as
the wall passes through. In order to fully understand what will happen
to the bubble, one may need to turn to numerical study, which shall
not be pursued here. In the following we will investigate three limits,
where the fate of the bubble can easily be predicted, and we believe
they together provide a more or less complete picture of black hole formation in our model. We found that the resulting black holes in the first limit have a mass spectrum with the most interesting features, so we shall mainly focus on this limit from the next section. The reader may thus skip the discussion of the second and the third limits (subsections \ref{small_rho_b} and \ref{large_rho_b}) for the main results in this work. 

\subsection{\label{negligible_rho_b}Negligible $\rho_{b}$ $\left(\rho_{b}\approx 0 \right)$}

\begin{figure}
\includegraphics[scale=0.61]{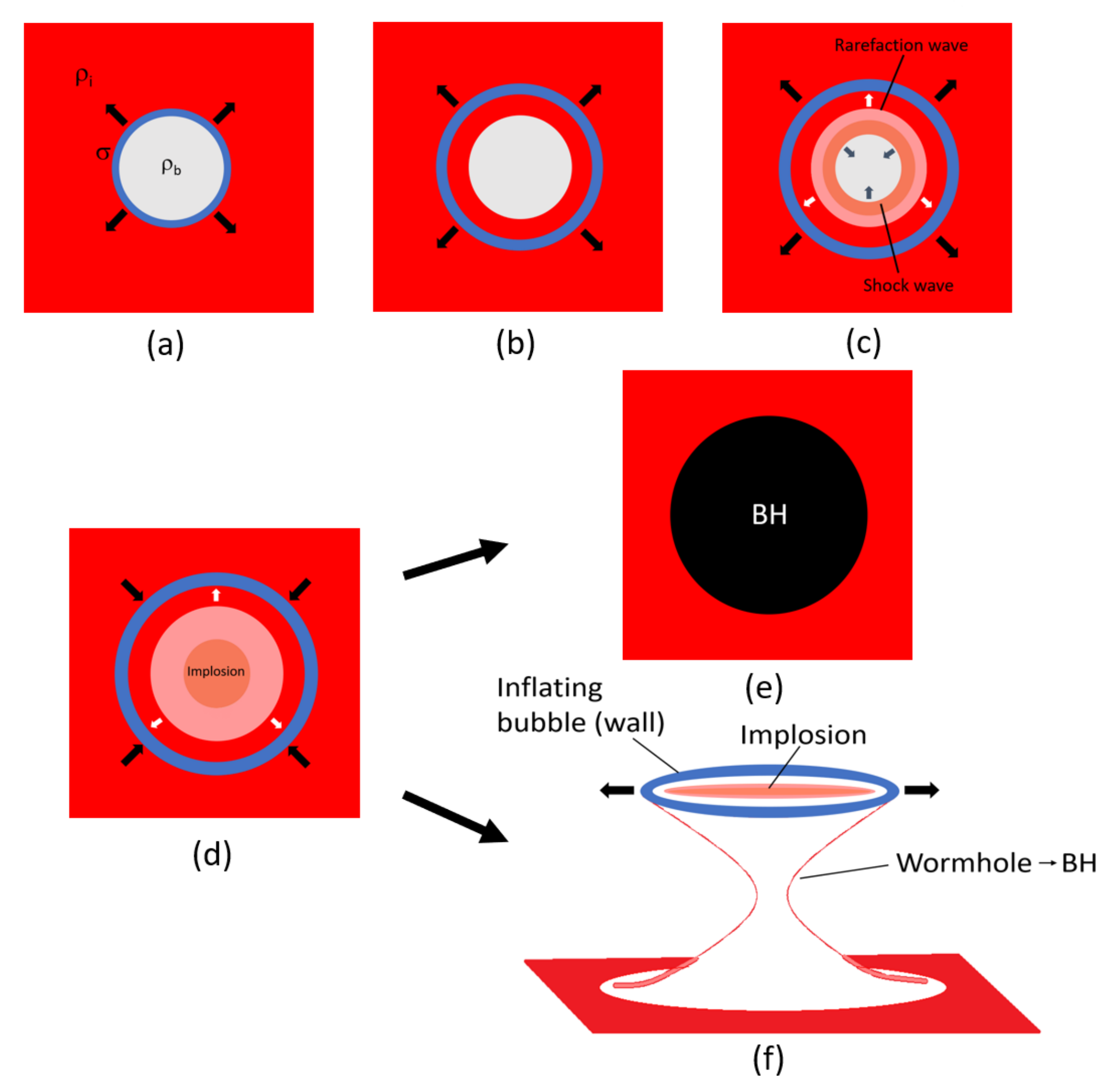}\caption{\label{fig:limit3_sketch}Cartoon pictures illustrating bubble evolution and black hole formation in the limit of negligible
$\rho_{b}$. (a) At the end of inflation, the exterior inflatons turn
into radiation fluid with density $\rho_{i}$. The bubble runs into
the fluid with a large Lorentz factor. (b) fluid freely flows inside
as the bubble expands into the homogeneous background. (c) Part of
the fluid inside the bubble forms a shock wave propagating inwards; part of the fluid forms a rarefaction wave that propagates
outwards, following the expanding bubble. (d) The shock wave implodes at the bubble center, which
may turn into a black hole or a reflected overdense wave. For an exterior observer, the bubble begins to turn around at $t_s$. (e) The bubble shrinks and collapses into a black hole
(subcritical), regardless what happens inside the bubble. (f) If $H_{\sigma}^{-1}\ll H^{-1}$,
the bubble wall would inflate and grow into a baby universe. This
creates a wormhole outside the bubble. The wormhole would eventually
close up and turn into a black hole (supercritical). The white regions inside and ``outside'' the bubble represent two almost empty shells produced by the repulsive domain wall. For an exterior FRW observer, there is a spherical rarefaction wave propagating outwards.}
\end{figure}

For the first limit, we assume the interior vacuum density
$\rho_{b}$ to be so small that the vacuum's gravitational effect is completely
negligible. As the bubble expands into the homogeneous radiation,
the ambient fluid flows into the interior vacuum, which is now essentially Minkowski. Part of the fluid forms a spherical shock wave that propagates inwards at the speed of sound, which then implodes at the bubble center. It is possible that, due to the massive accretion at the center, a black hole would form during the implosion. If there is not enough
fluid to form a black hole, the wave would get reflected and
damped as it propagates outwards as an overdense wave. Both phenomena are interesting on their own, but we do not intend to investigate them here, since the
implosion is not expected to affect the evolution of the bubble, which
takes place far from the central region. On the other hand, part of
the fluid inside the bubble, instead of going to the center, propagates
outwards (following the bubble wall) as a rarefaction wave, also at
the speed of sound. These processes are schematically shown in fig.
\ref{fig:limit3_sketch}. 

Since the interior vacuum is negligible, the bubble can be treated
as a spherical domain wall moving in the background of homogeneous
radiation, much like the case we considered in ref. \cite{Deng:2016vzb}, where the domain wall is comoving with the expanding universe after inflation, and behaves like a test wall. The main difference here is that the wall comes to a stop with respect to the Hubble flow at $t_{s}$ instead of $t_{i}$. 

After $t_{s}$, the comoving
radius of the bubble begins to decrease. A small bubble would later
reenter the Hubble horizon, shrink due to the wall tension, and eventually collapse into a black hole. This is what we call the ``subcritical'' regime. The resulting black hole mass is estimated to be the energy on the bubble wall when the bubble reaches its maximum physical size: $M\sim4\pi\sigma R_{max}^{2}$ \cite{Deng:2016vzb}. 

Furthermore, if the background radiation is homogeneous before the bubble size reaches $R_{max}$, since
the bubble collapses soon after Hubble crossing, $R_{max}$ can be
approximated by the Hubble horizon $R_{H}$ a test wall would fall
within (in an undisturbed FRW universe). We have $R_{H}= a(t_{H})\chi_{s}=\left(t_{H}/t_{i}\right)^{1/2}\chi_{s}$, where $t_H=R_H/2$ is the time at Hubble crossing. This then gives $R_{max}\sim R_{H}= H_{i}\chi_{s}^{2}$. 

However, this result is only valid if $R_{max}$ (or $R_H$) is reached before the rarefaction wave inside the bubble catches up with the bubble (see fig. \ref{fig:limit3_sketch}(d)(e)). The wave moves at the speed of sound, which is more than half of the expansion speed of the Hubble horizon, so it is possible that the bubble and the rarefaction wave meet before $t_H$. To be more specific, let $\chi_{w}(t)$ be the comoving radius of the rarefaction wave front, which satisfies $a\dot{\chi}_{w}=c_{s}$ (where $c_{s}=1/\sqrt{3}$ is the speed of sound) 
with initial condition $\chi_{w}(t_{i})=\chi_{i}$. Then we have
\begin{equation}
\chi_{w}(t)=\chi_{i}+2c_{s}\sqrt{t_{i}}\left(\sqrt{t}-\sqrt{t_{i}}\right).\label{wavefront}
\end{equation}
When it catches up with the test wall at $t$, we have $\chi_{w}(t)=\chi_s$,
which gives $t\sim H_i \left(\chi_s-\chi_i\right)^2$. This would be smaller than $t_H \sim H_i \chi_s^2$ if $\chi_s \sim \chi_i$, in which case the bubble's comoving radius at $t_i$ is so large that it does not grow much before $t_s$. Therefore, in order to use the result $R_{max} \sim R_H$, we need $\chi_s - \chi_i \sim a_s H_i^{-1}$ to be at least comparable to $\chi_i$ for all subcritical bubbles. In the following we assume this condition is satisfied, and leave more discussion to section \ref{discussion}. Therefore, the mass of the black hole from subcritical bubbles is estimated to be $M\sim4\pi\sigma R_H^{2}\sim H_{\sigma}H_{i}^{2}\chi_{s}^{4}$.

On the other hand, it is known that a planar domain wall tends to
inflate on the wall surface with a constant rate $H_{\sigma}$ \cite{Vilenkin:1984hy,Ipser:1983db}. If
the time scale $H_{\sigma}^{-1}$ is much smaller than $2t_{s}$ (the
Hubble time at $t_{s}$), which means the time is takes for the bubble
size to double is much shorter than the time it takes for the size
of the universe to double, the bubble would inflate. A wormhole would be created outside as the inflating bubble is surrounded by non-inflating radiation. The bubble grows without bound in a baby universe, which is connected to us by the wormhole throat. The wormhole throat would eventually pinch off, turning into a black hole (fig. \ref{fig:limit3_sketch}(f)).
This is what we call the ``supercritical'' regime. Due to the repulsive nature of the domain wall, the wall would be sandwiched by two almost empty layers (white regions in fig. \ref{fig:limit3_sketch}(f)). For an FRW
observer, the bubble wall turns around after $t_{s}$ and some fluid
with lower density flows out. What is left in the exterior universe
is a rarefaction wave propagating outwards into the homogeneous fluid at the speed of sound\footnote{Note that this is not the rarefaction wave inside the bubble.}. It is suggested by the simulations in ref. \cite{Deng:2016vzb} that the
black hole mass would be comparable to the horizon mass surrounded
by the wave front of the rarefaction wave at its Hubble crossing.
Let the comoving radius of the wave front be $\chi_{w}(t)$, which
satisfies $a\dot{\chi}_{w}=c_{s}$ with initial condition $\chi_{w}(t_{s})=\chi_{s}$.
The solution is
\begin{equation}
\chi_{w}(t)=\chi_{s}+2c_{s}\sqrt{t_{i}}\left(\sqrt{t}-\sqrt{t_{s}}\right).
\end{equation}
At Hubble crossing, we have $a(t)\chi_{w}(t)=2t$, which gives the
horizon mass
\begin{equation}
M\sim H_{i}\left(\chi_{s}-c_{s}a_{s}H_{i}^{-1}\right)^{2}.
\end{equation}
This is the estimate of the black hole mass for supercritical bubbles.
Note that the value of $a_{s}$ for a certain $\chi_{i}$ is determined
by solving eq. (\ref{chiEOM}). 

In conclusion, we can roughly approximate the black hole masses in
the limit of negligible $\rho_{b}$ by\footnote{Strictly speaking, there should be a prefactor $\sim\mathcal{O}(1\mbox{--}10)$ in eq. (\ref{massfunction-1}) from omitted prefactors in the mass estimates,
as well as from mass accretion after black hole formation, but this
does not affect our results by much.}
\begin{equation}
M\sim\begin{cases}
H_{\sigma}H_{i}^{2}\chi_{s}^{4}, & M_{min}<M<M_{*}\\
H_{i}\left(\chi_{s}-c_{s}a_{s}H_{i}^{-1}\right)^{2}, & M>M_{*}
\end{cases},\label{massfunction-1}
\end{equation}
where the minimum mass $M_{min}$ is from the fact that only
bubbles with $\chi_{i}>H_{i}^{-1}$ would collapse into black holes;
smaller ones are inside the Hubble horizon as they grow, and would
eventually shrink and disappear. The transition mass $M_{*}$ in eq.
(\ref{massfunction-1}) characterizes the scale that connects the
subcritical and supercrtical regimes. Note that $\chi_{s}-c_{s}a_{s}H_{i}^{-1}\sim\chi_{i}+\left(1-c_{s}\right)a_{s}H_{i}^{-1}\sim\chi_{s}$,
so the supercritical black hole mass is approximately $M\sim H_{i}\chi_{s}^{2}$.
Then the transition mass can be given by
\begin{equation}
M_{*}\sim H_{\sigma}^{-1},
\end{equation}
which corresponds to a bubble with radius
\begin{equation}
\chi_{*}\sim\left(\frac{H_{i}}{H_{\sigma}}\right)^{1/2}H_{i}^{-1}\label{critical_radius}
\end{equation}
at $t_{s}$. For such a critical bubble, the radiation density at
Hubble crossing is $\rho_{r}(t_{H})=\rho_{i}\left(t_{i}/t_{H}\right)^{2}\sim\rho_{i}\left(t_{i}/\chi_{*}\right)^{4}\sim H_{\sigma}^{2}$,
which is much larger than $\rho_{b}$ if $H_{\sigma}\gg H_{b}$. Hence
the gravitational effect of the interior vacuum can indeed be neglected,
and subcritical bubbles can be regarded as domain walls living in
an FRW background before entering the horizon. Moreover, when a subcritical
bubble forms a black hole, the contribution from the interior vacuum
to the mass is $\sim H_{b}^{2}M^{3}<\left(H_{b}M_{*}\right)^{2}M\sim\left(H_{b}/H_{\sigma}\right)^{2}M\ll M$
if $H_{b}\ll H_{\sigma}$. This means the black hole mass mainly comes
from the (kinetic) energy of the bubble wall, which is consistent
with the subcritical scenario discussed before.

\subsection{\label{small_rho_b}Small $\rho_{b}$ $\left( \rho_{b} \ll\rho_{r}(t_{s})\right)$}

For the second limit, we shall consider a $\rho_{b}$ much smaller
than the exterior radiation density $\rho_{r}$ when the bubble comes
to a stop with respect to the Hubble fluid, i.e., $\rho_{b}\ll\rho_{r}(t_{s})$.
This is the case when, for instance, the bubble wall tension $\sigma$
does not change much after inflation: by eq. (\ref{eq:a_s_limit2}),
we have $\rho_{b} \lesssim \rho_{i} a_{s}^{-5}$; on the other hand, due to Hubble expansion, we have $\rho_{i}=\rho_{r}(t_{s})a_{s}^{4}$;
hence $\rho_{b}\lesssim \rho_{i} a_{s}^{-5}=\rho_{r}(t_{s})a_{s}^{-1} \ll\rho_{r}(t_{s})$ (the last step is from our assumption that $a_s \gg 1$).
This implies the interior spacetime near the bubble wall can still
be approximated by an FRW universe dominated by radiation before $t_{s}$.
After $t_{s}$, the bubble's comoving radius begins to decrease but
its physical size continues to grow. During this course, the radiation
density inside the bubble decreases and may become comparable to $\rho_{b}$,
then the bubble interior (near the wall) is dominated by radiation
and a vacuum energy. 

A small bubble, after reaching its maximum size, would shrink
and collapse into a black hole. We expect the black hole mass to be comparable
to the mass of the bubble (interior vacuum energy plus bubble wall
energy) when the bubble size reaches the maximum $R_{max}$. After $t_{s}$, the bubble wall's trajectory is governed by (eq. (\ref{chiEOM2-1}))
\begin{equation}
\ddot{\chi}+\left(4-3a^{2}\dot{\chi}^{2}\right)H\dot{\chi}+\frac{2}{a^{2}\chi}\left(1-a^{2}\dot{\chi}^{2}\right)=-\left(\frac{\rho_{b}}{\sigma}-6\pi\sigma\right)\frac{\left(1-a^{2}\dot{\chi}^{2}\right)^{3/2}}{a}.\label{chiEOM_limit2}
\end{equation}
Here the scale factor and the Hubble parameter are solutions of the following Friedmann equations:
\begin{equation}
H^{2}=\frac{8\pi}{3}\left(\rho_{b}+\rho_{r}^{(in)}\right),\label{Friedmann_1_limit2}
\end{equation}
\begin{equation}
\frac{\ddot{a}}{a}=\frac{8\pi}{3}\left(\rho_{b}-\rho_{r}^{(in)}\right).\label{Friedmann_2_limit2}
\end{equation}
where $\rho_{r}^{(in)}(t)$ is the interior radiation density near
the bubble wall. The initial conditions are $a(t_{s})=a_{s}$ and $H(t_{s})\approx1/2t_{s}$. Then we can solve
eq. (\ref{chiEOM_limit2}) with initial conditions $\chi(t_{s})=\chi_{s}$
and $\dot{\chi}(t_{s})\approx0$. Then the mass of the black hole from bubble collapse is estimated to be
\begin{equation}
M\sim\frac{4}{3}\pi\rho_{b}R_{max}^{3}+4\pi\sigma R_{max}^{2}-8\pi\sigma^{2}R_{max}^{3},\label{eq:limit2_M_sub}
\end{equation}
independent of radiation. Here the first term is the interior vacuum
energy, the second is the bubble wall's surface energy, and the third
is the wall's binding energy. It is shown in the appendix that there
is a more precise way to find $M$, but the above equation gives a
good estimate.

If during the course of bubble growth $\rho_{b}$ becomes comparable
to $\rho_{r}^{(in)}$ and the bubble size exceeds $\sim H_{b}^{-1}$,
the bubble would inflate. Similar to the supercitical scenario discussed in the last subsection, the inflating bubble grows in a baby universe connected to our universe by a wormhole, which would later close off and turn into a black hole. Ref. \cite{Deng:2017uwc} suggests that the mass of the black hole from such a bubble can also be estimated to be $M\sim H_{i}\left(\chi_{s}-c_{s}a_{s}H_{i}^{-1}\right)^{2}$, the same as that from the last subsection.

\subsection{\label{large_rho_b}Large $\rho_{b}$ $\left(\rho_{b}\gtrsim\rho_{r}(t_{s})\right)$}

\begin{figure}
\includegraphics[scale=0.45]{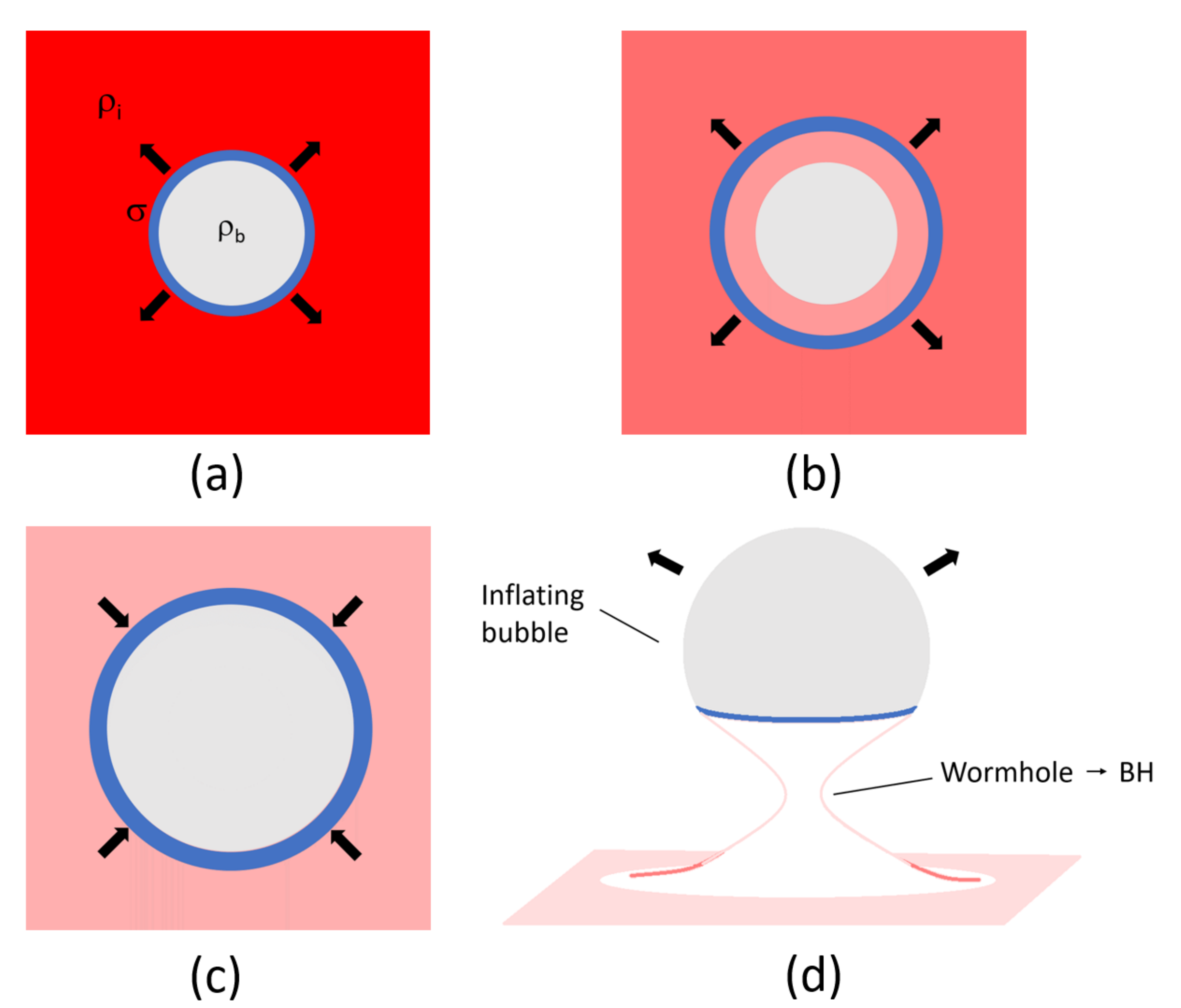}\caption{\label{fig:limit1_sketch}Cartoon pictures illustrating  bubble evolution and black hole formation in the limit of very large
$\rho_{b}$. (a) After inflation, the bubble runs into
the fluid with a large Lorentz factor. (b) fluid freely flows inside as the
bubble expands into the homogeneous background. Due to the large $\rho_b$, fluid that flows in soon gets diluted. (c)
As the radiation density outside the bubble decreases due to Hubble
expansion, fluid that flows into the growing bubble becomes insignificant,
and the bubble mass is dominated by the interior vacuum and the bubble
wall. For an exterior observer, the bubble begins to turn around at $t_s$. (d) After $t_s$, the physical size of the bubble continues to grow and then exceeds the de Sitter horizon associated
with the interior vacuum. The bubble would then inflate into a baby universe,
which is connected to us by a wormhole. The wormhole would eventually
pinch off, turning into a black hole. The white region around the wormhole throat represents an almost empty shell, where the fluid gets significantly diluted by the bubble's interior vacuum before $t_s$. For an exterior FRW observer, there is a spherical rarefaction wave propagating outwards.}
\end{figure}

The exterior radiation density is initially larger than the interior
vacuum density ($\rho_{i}>\rho_{b}$), but it gets diluted by the
cosmic expansion as the bubble grows. For the third
limit, we assume $\rho_{b}$ to be sufficiently large so that it is
comparable to or much larger than the exterior radiation density $\rho_{r}(t)$ at $t_{s}$, i.e., $\rho_{b}\gtrsim\rho_{r}(t_{s})$. By the definition of $a_{s}$, we have
\begin{equation}
\rho_{i}a_{s}^{-4}=\rho_{r}(t_{s})\lesssim\rho_{b}.
\end{equation}
Then the bubble radius at $t_{s}$ satisfies (eq. (\ref{R_s}))
\begin{equation}
R_{s}\sim a_{s}\chi_{i}+a_{s}^{2}H_{i}^{-1}>a_{s}^{2}H_{i}^{-1}=\left(\frac{8\pi}{3}\rho_{i}a_{s}^{-4}\right)^{-1/2}\gtrsim\left(\frac{8\pi}{3}\rho_{b}\right)^{-1/2}=H_{b}^{-1}.
\end{equation}
Note that $H_{b}^{-1}$ is the de Sitter horizon associated with the
interior vacuum. Since $\rho_{b}\gtrsim\rho_{r}(t_{s})$, $R_{s}\gtrsim H_{b}^{-1}$ implies all bubbles would inflate in this limit (fig. \ref{fig:limit1_sketch}), and the resulting black hole masses are give by $M\sim H_{i}\left(\chi_{s}-c_{s}a_{s}H_{i}^{-1}\right)^{2}$. 

\section{\label{Mass spectrum} Mass spectrum}

In the previous sections we have shown how a black hole could be formed
by a vacuum bubble and estimated the black hole mass as a function
of the bubble radius at $t_{s}$ in three limits. During inflation,
bubbles grow exponentially  and their sizes
spread over a large range of scales, therefore the resulting black
holes should have an extended mass spectrum. Now we would like to
compute this spectrum.

Bubbles formed earlier expand to larger sizes, but their number density
gets more diluted by the cosmic expansion. By assuming that the bubbles
are formed with a vanishing size and that the bubble worldsheet can
be approximated by the future light cone of the nucleation point,
one can find that the number density of bubbles having radius in the
interval $(\chi_{i},\chi_{i}+d\chi_{i})$ at $t$ ($t>t_{i}$) is \cite{Deng:2017uwc}
\begin{equation}
dn(t)\approx\lambda\frac{d\chi_{i}}{a(t)^{3}\left(\chi_{i}+H_{i}^{-1}\right)^{4}}.
\end{equation}
where $\lambda$ is the bubble nucleation rate per Hubble spacetime volume $H_i^{-4}$. 

It is convenient to use the standard mass function to characterize
the spectrum of PBHs,
\begin{equation}
f(M)=\frac{M^{2}}{\rho_{CDM}(t)}\frac{dn(t)}{dM},
\end{equation}
where $\rho_{CMD}$ is the mass density of cold dark matter. This
can be interpreted as the fraction of dark matter in PBHs at $M$ within the mass
range $\Delta M\sim M$. Because the black hole density and dark matter
density are diluted by the cosmic expansion in the same way, $f(M)$
is a constant over time. The total fraction of dark matter in PBHs
can be obtained by integrating the mass function, and is given by
\begin{equation}
f_{PBH}=\int\frac{dM}{M}f(M),
\end{equation}
which should satisfy $f_{PBH}\leq1$. During the radiation era,
\begin{equation}
\rho_{CDM}(t)\sim\frac{M_{Pl}^{3}}{Bt^{3/2}\mathcal{M}_{eq}^{1/2}},
\end{equation}
where $B\sim10$ is a constant and $\mathcal{M}_{eq}\sim10^{17}M_{\odot}$
is the dark matter mass within a Hubble radius at dust-radiation equality.
Then the mass function can be written as
\begin{equation}
f(M)\sim\frac{B\lambda\mathcal{M}_{eq}^{1/2}}{M_{Pl}^{3}}\frac{M^{2}}{H_{i}^{3/2}\left(\chi_{i}+H_{i}^{-1}\right)^{4}}\frac{d\chi_{i}}{dM}.\label{f(M)}
\end{equation}

In the following we will only show results in the limit of negligible
$\rho_{b}$, because we found that mass spectra in this limit have
the most interesting features in our model. In order to find
$f(M)$, we need the relation between $M$ and $\chi_{i}$. This can
be obtained by numerically solving eq. (\ref{chiEOM}) and using eq.
(\ref{massfunction-1}). The mass function is completely determined
by the following five quantities: the Lorentz factor of the bubble
wall $\gamma_{i}$ at the end of inflation, the bubble wall tension
$\sigma$, the interior vacuum density $\rho_{b}$, the inflationary
density $\rho_{i}$, and the bubble nucleation rate $\lambda$. It
would be more convenient to specify the mass function with parameters
$\gamma_{i},\lambda,M_{*}\equiv H_{\sigma}^{-1}$ and $\eta_{i}\equiv H_{i}^{1/2}$, where $\eta_i$ characterizes the inflationary energy scale. The specific value
of $\rho_{b}$ is not important as long as we have $H_{b}\ll H_{\sigma}$.

\begin{figure}
\includegraphics[scale=0.4]{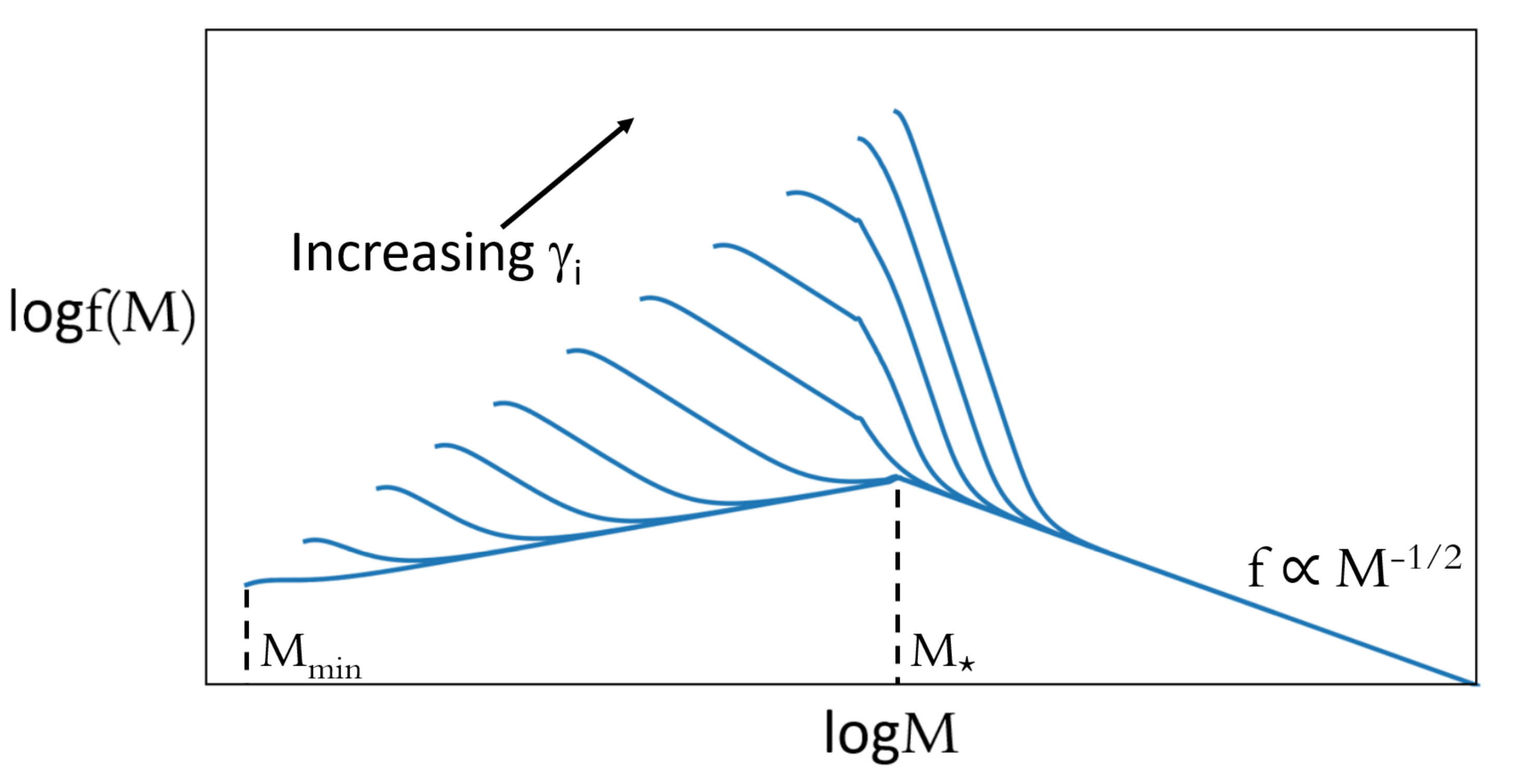}

\caption{\label{fig:limit1+2}Some examples of the mass function $f(M)$ in the limit of negligible $\rho_{b}$, covering masses above $M_{min}$. We have fixed parameters except for the Lorentz factor $\gamma_{i}$. For small $\gamma_{i}$ the maximum of $f(M)$ appears at $M_{*}$.
As $\gamma_{i}$ increases, another peak develops at $M_{min}$. For
large $\gamma_{i}$, $f(M_{min})$ becomes larger than $f(M_{*})$,
and the spectrum turns spiky. For large masses, all spectra approach
to $f\propto M^{-1/2}$.}
\end{figure}

Several curves of $f(M)$ are shown in fig. \ref{fig:limit1+2}. Depending
on the parameters, the mass function can have very different shapes
and can be either wide or relatively narrow. Specifically, in fig. \ref{fig:limit1+2} we
fix all other parameters except for the Lorentz factor $\gamma_{i}$.
For small $\gamma_{i}$, the maximum of $f(M)$ appears at $M_{*}$, and $f(M)\propto M^{1/4}$ near $M_*$ in the subcritical regime. This is consistent with the results in refs. \cite{Garriga:2015fdk, Deng:2016vzb}.
As $\gamma_{i}$ increases, another peak develops at $M_{min}$ (which is the black hole mass from a bubble with $\chi_i = H_i^{-1}$), and the spectrum becomes gradually spiky, with the black hole
population dominated by those with mass $\sim M_{min}$. When $\gamma_{i}$ is sufficiently large, all bubble are in the supercritical regime.
We can also see that for large masses, all spectra converge to the
same straight line. It is shown in the appendix that for very large
bubbles ($\chi_{i}\gg H_{i}^{-1}$), $a_{s}$ is a constant independent
of $\chi_{i}$ (eq. (\ref{a_s})). For supercritical bubbles, we have $M\sim H_{i}\chi_{s}^{2}\sim H_{i}\left(\chi_{i}+a_{s}H_{i}^{-1}\right)^{2}$,
then by eq. (\ref{f(M)}) we have
\begin{equation}
f(M)\propto\frac{M^{-1/2}}{\left[1-a_{s}\left(H_{i}M\right)^{-1/2}\right]^{4}}.
\end{equation}
For $M\to\infty$, this gives $f(M)\propto M^{-1/2}$, which explains
the converging behavior of the mass functions for large masses. We also note that the values of $f(M_{min})$ in different spectra (of varying $\gamma_i$) seem to be on a straight line in fig. \ref{fig:limit1+2}, which suggests a power function $f(M_{min})\propto M_{min}^\alpha$. Since the  explicit relation between $M$ and $\chi_i$ (hence the relation between $f(M)$ and $M$) is unknown, the power $\alpha$ cannot be determined analytically. Numerically we found $\alpha \approx 1$. 

\section{\label{Observational bounds} Observational bounds}

The mass function of a certain PBH model is constrained by observations:
based on dynamical, microlensing and astrophysical effects, different
observational missions have placed (stringent) upper bounds on the
population of PBHs in different mass ranges. Some conservative and
up-to-date constraints are shown in fig. \ref{fig:f_max} on the total
fraction of dark matter in PBHs at $\sim10^{15}\mbox{--}10^{47}\ \text{g}$.
The colored regions have been excluded by observations. These bounds
are valid only for monochromatic PBHs, i.e., PBHs of almost the same
mass with a narrow spectrum. We can see that the only window allowing
PBHs to constitute all dark matter is at $\sim10^{17}\mbox{--}10^{23}\ \text{g}$.
If we would like to explain LIGO/Virgo events with PBHs, there should
be a sufficient number of them around $\sim 10M_{\odot}$. Likewise,
if SMBHs were seeded by PBHs, there should be enough of them at $M\gtrsim10^{3}M_{\odot}$
such that the number density of the seeds
exceeds that of galaxies observed today. With an increasing number
of observations and studies imposing more and more bounds, it seems
difficult for PBHs to completely account for all three of them, unless
the mass spectrum has multiple spikes \cite{Cai:2018tuh,Carr:2018poi}.

\begin{figure}
\includegraphics[scale=0.33]{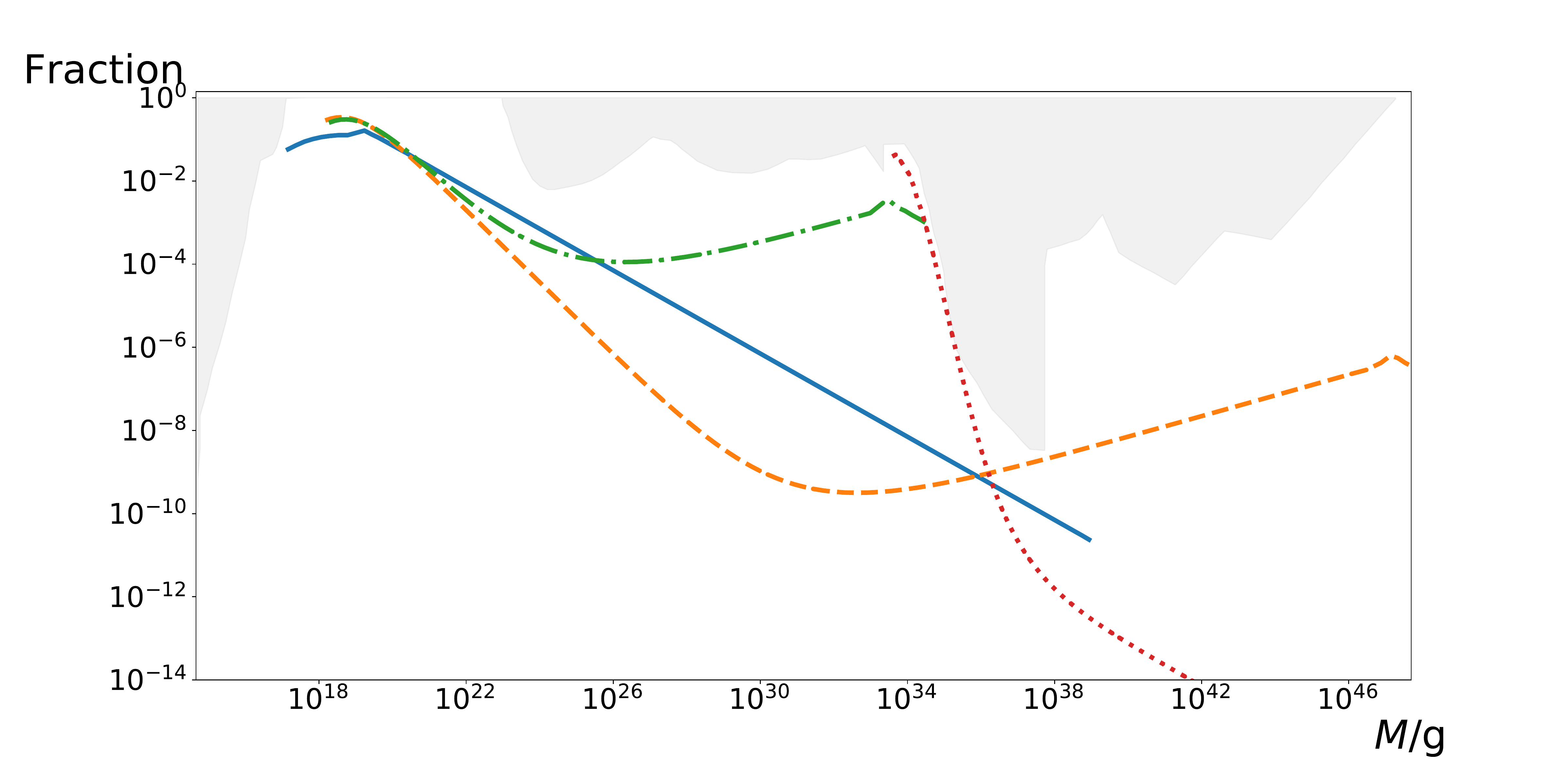}

\caption{\label{fig:limit1}The light grey shaded areas are observationally
excluded regions (also shown in fig. \ref{fig:f_max}). Their bottom
boundary is $f_{max}(M)$, the upper bound of the fraction of dark
matter in PBHs for a monochromatic mass distribution. The colored
curves are four examples of $f(M)$ in the limit of negligible $\rho_{b}$,
where $f(M)$ can be interpreted as the fraction of dark matter in
PBHs within the mass range $M\mbox{--}2M$ for an extended mass distribution.
$f_{max}(M)$ and $f(M)$ represent different quantities but we place
them in the same figure because $f(M)$ can loosely be constrained
by $f_{max}(M)$. The mass range $10^{17}\mbox{--}10^{23}\ \text{g}$
is poorly constrained by observations at the moment, so $f(M)=\mathcal{O}(0.1\mbox{--}1)$
in this window could provide an explanation to all dark matter. The
blue (solid) and orange (dashed) curves are two spectra where the PBHs
can account for all dark matter and seeds for SMBHs; the green (dashdotted)
curve can explain both dark matter and LIGO/Virgo black holes; the red
(dotted) curve can explain LIGO/Virgo black holes and SMBHs.}
\end{figure}

For an extended mass spectrum, such as the one given by eq. (\ref{f(M)}),
fig. \ref{fig:f_max} cannot be adopted directly. But one can apply
the method developed in ref. \cite{Carr:2017jsz}, using the bounds in fig. \ref{fig:f_max}
to constrain the parameters in the extended spectrum. Ref. \cite{Carr:2017jsz}
shows that the mass function $f(M)$ should satisfy the following condition,
\begin{equation}
\sum_{j}\left(\int_{M_{j_{1}}}^{M_{j_{2}}}\frac{dM}{M}\frac{f(M)}{f_{j,max}(M)}\right)^{2}<1,
\end{equation}
where $j$ denotes different observations, with $\left(M_{j_{1}},M_{j_{2}}\right)$ the mass range covered by that particular observation, and $f_{j,max}(M)$ the corresponding upper bound. Loosely speaking, however, as
long as $f(M)/M$ does not have a plateau over a relatively big range,
we can constrain a certain model by placing $f(M)$ in fig. \ref{fig:f_max}
while avoiding hitting the excluded regions. For this reason, we show
$f(M)$ and $f_{max}(M)$ in the same figure (fig. \ref{fig:limit1}),
even though they represent different quantities. 

Fig. \ref{fig:limit1} shows four interesting examples of our mass
function in the limit of negligible $\rho_{b}$. Firstly, the blue
(solid) curve and the orange (dashed) curve are two examples consistent
with PBHs being all dark matter as well as the seeds for SMBHs. The
blue one is produced by setting $\gamma_{i}=10^{3},M_{*}=10^{20}\ \text{g},\lambda\approx10^{-17}$
and $\eta_{i}\approx10^{8}\ \text{GeV}$, and covers the mass range
beyond $\sim10^{17}\ \text{g}$; the orange one is from $\gamma_{i}=10^{23},M_{*}=10^{20}\ \text{g},\lambda\approx10^{-9}$
and $\eta_{i}\approx10^{5}\ \text{GeV}$, and covers the mass range
beyond $\sim10^{18}\ \text{g}$. Both curves have their maximum value
$f=\mathcal{O}(0.1\mbox{--}1)$ on the lower end, where $f_{max}(M)$
is poorly constrained. As a result, PBHs with these
spectra can account for all dark matter ($f_{PBH}=1$). In addition,
these two spectra may also provide the seeds for SMBHs. At the present
time ($t_{0}$), the number density of PBHs of mass $\sim M$ is approximately given by
\begin{equation}
n(M)\sim\rho_{CDM}(t_{0})\frac{f(M)}{M}\approx4\times10^{11}\left(\frac{M_{\odot}}{M}\right)f(M)\ \text{Mpc}^{-3}.
\end{equation}
It was shown in ref. \cite{Serpico:2020ehh} that PBHs at $z\sim6$ with masses $10^{2}M_{\odot}\lesssim M\lesssim10^{4}M_{\odot}$
can attain sufficient accretion, growing to SMBHs at the present time,
even if $f(M)$ is as small as $10^{-9}$ in that mass range. Therefore,
in order to serve as seeds for SMBHs, PBHs should have masses $M\gtrsim10^{2}M_{\odot}$.
We can see from the two curves that $f(10^{3}M_{\odot})\sim10^{-9}$,
which gives $n(10^{3}M_{\odot})\sim0.4\ \text{Mpc}^{-3}$. This
is larger than the galaxy density $n_{G}\sim0.1\ \text{Mpc}^{-3}$.
Therefore it is possible that SMBHs observed today were seeded by
these PBHs.

The green (dashdotted) curve in fig. \ref{fig:limit1} is produced by setting $\gamma_{i}=10^{11},M_{*}=M_{\odot},\lambda\approx10^{-12}$
and $\eta_{i}\approx10^{6}\ \text{GeV}$. It covers the mass range
$\sim10^{18}\mbox{--}10^{35}\ \text{g}$. PBHs with this spectrum can
also account for all dark matter. As the mass increases, $f(M)$ reaches
a minimum at $\sim10^{26}\ \text{g}$ and increases to $f\sim0.01$
around the transition mass $M_{*}=M_{\odot}$. In order to reproduce the black hole merger rate inferred by the LIGO/Virgo events, the mass function should have $f(M\sim 10M_{\odot})=\mathcal{O}(10^{-3}\mbox{--}10^{-2})$ \cite{Sasaki:2016jop,Ali-Haimoud:2017rtz,Vaskonen:2019jpv,DeLuca:2020qqa}. 
Therefore, our model is able to provide a primordial explanation to
those detections. However, note that in order not to hit the bound imposed
by black hole accretion \cite{Serpico:2020ehh} (the pink region in fig. \ref{fig:f_max}), there should be a cutoff at around $200M_{\odot}$.
Physically, such a cutoff can be achieved if we relax the assumption
that bubbles nucleate with a constant rate during inflation. This
is possible especially in inflationary models where the inflaton travels
for a relatively large distance during the slow-roll. 

Lastly, the red (dotted) curve in fig. \ref{fig:limit1} is $f(M)$
with $\gamma_{i}=10^{18},M_{*}=10M_{\odot},\lambda=10^{-19}$ and
$\eta_{i}\approx1\ \text{TeV}$. It covers the mass range above $M_\odot$.
With this spectrum, we have $f(10M_{\odot})=\mathcal{O}(10^{-3}\mbox{--}10^{-2})$,
hence these PBHs can account for the LIGO/Virgo black holes. In addition,
we can see that $f(10^{3}M_{\odot})\sim10^{-9}$, so these PBHs
could also be the seeds of SMBHs. The maximum of mass function is
at $M_{min}\sim M_{\odot}$, where $f=\mathcal{O}(0.01\mbox{--}0.1)$, so
theses black holes can constitute no more than $10\%$ of the dark
matter. Another noticeable feature of this spectrum is that it corresponds
to an inflationary energy scale $\eta_{i}\sim1\ \text{TeV},$ which
is the smallest one can envision for inflation.

\section{\label{discussion} Summary and discussion}

In this paper we have studied a mechanism of primordial black hole
formation, where the black holes are formed by vacuum bubbles that
randomly nucleate during inflation. This is a natural outcome if the inflaton
runs in a general, multi-dimensional potential, where there could
be many vacua and thus many tunneling channels. Typically, these bubbles
expand with an acceleration and soon acquire a large Lorentz factor.
Inflatons outside the bubbles thermalize into radiation fluid at the
end of inflation, which is then run into by the fast expanding bubbles.
If there are no couplings between the bubble fields and the standard model particles other
than gravity, the fluid would then freely flow in as the bubbles grow
possibly by many orders of magnitude. In the meantime the interior
vacuum and the wall tension tend to slow down the bubble expansion.
The bubble wall's trajectory can be found by matching the spacetimes
on two sides of the wall without the knowledge of the fluid's motion,
since the exterior spacetime can be described by a flat FRW metric. This
works till the time when the bubble comes to a halt with respect to
the Hubble flow (at a comoving radius denoted by $\chi_s$), because after that the bubble turns around and interacts with the interior inhomogeneous fluid, and neither side of the wall can now be expressed in a closed form. Therefore, in order to accurately describe the evolution of the bubbles as well as the radiation fluid,
one needs to turn to numerical study. This may be challenging because
one has to deal with a thin wall with a large Lorentz factor traveling a long distance. 

However, if we focus on the case of a small bubble wall tension,
and consider the extreme limits of small and large interior vacuum density
$\rho_{b}$, the fate of the bubble can easily be estimated. In the
limit of very small $\rho_{b}$, where $\rho_{b}$ is assumed to be
at all times subdominant compared to the radiation density, the bubble
can be regarded as a spherical domain wall moving in homogeneous fluid,
since the bubble grows at a speed close to that of light for most
of the time before it comes to a stop. In the limit of very large $\rho_{b}$, the fluid that
flows in soon gets diluted and could eventually by neglected compared
to the interior vacuum, hence the bubble can be regarded as a pure
vacuum bubble. Black holes would form in both scenario, either in
the subcritical or in the supercritical regime, and we found estimates for
the resulting black hole masses.

For completeness, we shall now (qualitatively) comment on a situation not captured in our previous analysis (subsection \ref{negligible_rho_b}). In the limit of negligible $\rho_b$, the estimate of the black hole mass from subcritical bubbles is directly related to $R_{max}$, the maximum value of the bubble's physical radius, which can be approximated by the cosmic horizon at the Hubble crossing of a test wall at $\chi_s$. However, by doing this we have assumed that the interior spacetime near the bubble wall can be approximated by an FRW metric dominated by radiation (see fig. \ref{fig:limit3_sketch}), but this is no longer true if the bubble comes within the spherical rarefaction wave before Hubble crossing, which would happen to bubbles with $\chi_s\sim\chi_i$, as we found in subsection \ref{negligible_rho_b}. This can be avoided if $a_s \sim H_i \left(\chi_s-\chi_i\right)$ is sufficiently large for all subcritical bubbles, but if it did happen to some bubbles, possibly close to the transition regime, they may either (i) become supercritical because the region inside the rarefaction wave front has a lower radiation density, thus a smaller cosmic expansion rate, which may be smaller than the expansion rate $H_\sigma$ on the bubble wall; or (ii) acquire a $R_{max}<R_H$ due to the smaller cosmic expansion rate inside the wave front, in which case the estimate $M\sim 4\pi\sigma R_H^2$ becomes an upper bound. In either case, we expect that this would simply smooth the mass function near the transition regime, instead of bringing a spike.  

Our main results are in fig. \ref{fig:limit1}, where we show several
examples of the mass spectrum of the black holes in the limit of very small $\rho_b$. Depending
on the model parameters, the mass distribution of these black holes
could either be extended or spiky, and may have two peaks separated by a wide mass range.
Constrained by the current observational bounds, our model is able
to simultaneously explain the origin of any two of the following three
puzzles: dark matter, black holes detected by LIGO/Virgo ($M=\mathcal{O}(10\mbox{--}100)M_{\odot}$),
and supermassive black holes at the center of most galaxies ($M=\mathcal{O}(10^{6}\mbox{--}10^{10})M_{\odot}$).
Of course, one could envisage that there exist multiple tunneling
channels during inflation, which allow the formation of bubbles with
different properties. Then it is possible that the resulting black
holes have a spectrum with more than two peaks.

Finally, we note that all astrophysically interesting spectra in our
model require a small energy scale of inflation, the largest one being
$10^{8}\ \text{GeV}$, and the smallest one $1\ \text{TeV}$, which
is the lowest one would expect for an inflationary model (although the
definite bound is at the MeV scale, consistent with Big Bang Nucleosynthesis).
A lower bound on the inflationary scale possibly determined
in the future may thus impose a serious constraint on our model.

\section*{Appendix: Spacetime matching on bubble wall}

In this appendix we study the dynamics of the bubble by matching the spacetimes on two
sides of the wall according to the Israel junction conditions \cite{Israel:1966rt,Maeda:1985ye}. The bubble wall's equation of motion will be derived, as
well as the bubble mass during the collapse. We closely
follow the method and notations in ref. \cite{Tanahashi:2014sma}

After inflation, the bubble runs into the ambient radiation fluid
at a speed close to that of light. Since we assume there is only gravitational
interaction between the bubble and the fluid, the fluid freely flows
in, and the exterior spacetime can be described by a flat FRW metric,
\begin{equation}
ds^{2}=dt^{2}-a^{2}(t)(dr^{2}+r^{2}d\Omega^{2}).
\end{equation}
The metric of the bubble interior can be expressed as 
\begin{equation}
ds^{2}=dt^{2}-a_{1}(r,t)^{2}dr^{2}-a_{2}(r,t)^{2}r^{2}d\Omega^{2}.\label{metric2}
\end{equation}
For a thin wall, we can use the Israel junction conditions to match
these two metrics.

Let the trajectory of the wall be $(t(\tau),\chi(\tau)),$ where $\tau$
is the proper time on the wall. Then a vector tangent to the wall
is $v^{\mu}=(t_{,\tau},\chi_{,\tau})$, where $_{,\tau}\equiv d/d\tau.$
Assuming $\dot{t}$ to be positive and $v^{\mu}v_{\mu}=-1$, we have
$t_{,\tau}=\sqrt{1+a_{1}^{2}\chi_{,\tau}^{2}}.$ Let $\xi^{\mu}$
be the normal unit vector on the wall. Then $\xi^{\mu}v_{\mu}=0$
and $\xi^{\mu}\xi_{\mu}=1$ give $\xi^{\mu}=(a_{1}\chi_{,\tau},t_{,\tau}/a_{1})$
and $\xi_{\mu}=(-a_{1}\chi_{,\tau},a_{1}t_{,\tau})$. We also define
the notations $[Q]\equiv Q_{out}-Q_{in}$, $\{Q\}\equiv Q_{out}+Q_{in}$,
and $\bar{Q}\equiv(Q_{out}+Q_{in})/2$, where ``out'' and ``in''
denote matching quantities on different sides of the hypersurface. 

The induced metric on the wall is defined to be $h_{\mu\nu}=g_{\mu\nu}-\xi_{\mu}\xi_{\nu}$.
Then the first junction condition is $[h_{\mu\nu}]=0$. The $(t,t)$
and $(\theta,\theta)$ components give 
\begin{equation}
a_{1}^{2}=a_{2}^{2}=a^{2}.
\end{equation}
Therefore $a_{1,\tau}=a_{2,\tau}=a_{,\tau}$, which gives 
\begin{equation}
\dot{a}t_{,\tau}=a_{,\tau}=t_{,\tau}\dot{a_{1}}+r_{,\tau}a_{1}^{\prime}=t_{,\tau}\dot{a_{2}}+r_{,\tau}a_{2}^{\prime},
\end{equation}
where the overdot and the prime stand for the first derivative with
respect to $t$ and $r$, respectively.

The extrinsic curvature of the wall is $K_{\mu\nu}=h_{\mu}^{\ \alpha}\nabla_{\alpha}\xi_{\nu}$,
where $\nabla$ is the covariant derivative operator for 4-spacetime.
Then the second junction condition is $[K_{\mu\nu}]=8\pi(-S_{\mu\nu}+Sh_{\mu\nu}/2)$,
where $S_{\mu\nu}=-\sigma h_{\mu\nu}$ is the energy-momentum tensor
of the domain wall, $\sigma$ being the surface energy. The $(\theta,\theta)$
component then gives 
\begin{equation}
\left[\xi^{\mu}\partial_{\mu}\ln(a_{2}\chi)\right]=-4\pi\sigma.\label{junction1}
\end{equation}
Right inside and outside the bubble wall, we have 
\begin{equation}
\left.\xi^{\mu}\partial_{\mu}\ln\left(a_{2}\chi\right)\right|_{in}=a_{1}\chi_{,\tau}\frac{\dot{a_{2}}}{a_{2}}+\frac{t_{,\tau}}{a_{1}}\left(\frac{a_{2}^{\prime}}{a_{2}}+\frac{1}{\chi}\right),
\end{equation}
\begin{equation}
\left.\xi^{\mu}\partial_{\mu}\ln\left(a\chi\right)\right|_{out}=r_{,\tau}\dot{a}+\frac{t_{,\tau}}{a\chi},
\end{equation}
respectively. Then eq. (\ref{junction1}) yields 
\begin{equation}
a_{2}^{\prime}=4\pi\sigma a^{2}\sqrt{1+\left(a\chi_{,\tau}\right)^{2}}.
\end{equation}
The $(\tau,\tau)$ component of the second junction condition gives
\begin{equation}
\left[\xi_{\mu}D_{\tau}v^{\mu}\right]=-4\pi\sigma,\label{junction2}
\end{equation}
where $D_{\tau}v^{\mu}=\partial_{\tau}v^{\mu}+\Gamma_{\lambda\sigma}^{\mu}v^{\lambda}v^{\sigma},$
with $\Gamma_{\lambda\sigma}^{\mu}$ the Chistoffel symbols in 4-spacetime.
Right inside and outside the bubble wall, we have 
\begin{equation}
\left.\xi_{\mu}D_{\tau}v^{\mu}\right|_{in}=\frac{\left(a_{1}\chi_{,\tau}\right)_{,\tau}}{t_{,\tau}}+\dot{a_{1}}\chi_{,\tau},
\end{equation}
\begin{equation}
\left.\xi_{\mu}D_{\tau}v^{\mu}\right|_{out}=\frac{a\chi_{,\tau\tau}+2a_{1,\tau}\chi_{,\tau}}{t_{,\tau}},
\end{equation}
respectively. Hence eq. (\ref{junction2}) gives 
\begin{equation}
a_{1}^{\prime}=-\frac{4\pi\sigma\sqrt{1+\left(a\chi_{,\tau}\right)^{2}}}{\chi_{,\tau}^{2}}.
\end{equation}
Converting $\tau$ to $t$, $a_{1}^{\prime}$ and $a_{2}^{\prime}$
can be rewritten as 
\begin{equation}
a_{1}^{\prime}=-\frac{4\pi\sigma\sqrt{1-\left(a\dot{\chi}\right)^{2}}}{\dot{\chi}^{2}},\label{eq:a1prime}
\end{equation}
\begin{equation}
a_{2}^{\prime}=\frac{4\pi\sigma a^{2}}{\sqrt{1-\left(a\dot{\chi}\right)^{2}}}.\label{eq:a2prime}
\end{equation}

The equation of motion for the wall is given by $S_{\mu\nu}\bar{K}^{\mu\nu}=[T_{\mu\nu}\xi^{\mu}\xi^{\nu}].$
A perfect fluid has energy-momentum tensor $T_{\mu\nu}=(\rho+p)u_{\mu}u_{\nu}+pg_{\mu\nu}$.
Note that the interior vacuum pressure $p_{b}=-\rho_{b}$, while the
radiation pressure $p_{r}=\rho_{r}/3$. Then we have
\begin{equation}
\left\{ \xi_{\mu}D_{\tau}v^{\mu}+2\xi^{\mu}\partial_{\mu}\ln(a_{2}\chi)\right\} =-\frac{2}{\sigma}\left[(\rho+p)(u^{\mu}\xi_{\mu})^{2}+p\right].\label{eq:EOM}
\end{equation}
Now we assume $\left[u_{\mu}\xi^{\mu}\right]=0,$ and $\left[\rho_{r}\right]=0$.
Since $u_{\mu}\xi^{\mu}$ is the 4-velocity of the fluid in the direction
of the unit normal vector, this two conditions mean the radiation
fluid flows through the wall smoothly. Then the equation of motion
(eq. (\ref{eq:EOM})) yields
\begin{equation}
\ddot{\chi}+\left(4-3a^{2}\dot{\chi}^{2}\right)H\dot{\chi}+\frac{2}{a^{2}\chi}\left(1-a^{2}\dot{\chi}^{2}\right)=-\left(\frac{\rho_{b}}{\sigma}+6\pi\sigma\right)\frac{\left(1-a^{2}\dot{\chi}^{2}\right)^{3/2}}{a},\label{chiEOM2}
\end{equation}
where $H\equiv\dot{a}/a=(2t)^{-1}$ is the Hubble parameter. Let $t_{i}$
be the time when inflation ends, and the scale factor $a=\left(t/t_{i}\right)^{1/2}$.
The bubble expands with initial conditions $\chi(t_{i})=\chi_{i}$
and $\dot{\chi}(t_{i})=\left(1-\gamma_{i}^{-2}\right)^{1/2}$, where
$\gamma_{i}$ is the Lorentz factor of the bubble wall at $t_{i}$.
Since $\gamma_{i}$ is assumed to be large, the trajectory of the
wall can be approximated by $a\dot{\chi}\approx1$ (at least at early
times), where the overdot represents the first derivative with respect
to $t$. This gives
\begin{equation}
\chi(t)\approx\chi_{i}+2\sqrt{t_{i}}(\sqrt{t}-\sqrt{t_{i}}),\label{chi-1}
\end{equation}

For a bubble far beyond the Hubble horizon at $t_{i}$, the trajectory
of the wall can be obtained analytically. For an exterior FRW observer,
the Lorentz factor of the wall is $\gamma(t)=\left(1-a^{2}\dot{\chi}^{2}\right)^{-1/2}$.
Define $u\equiv\sqrt{\gamma^{2}-1}$, then eq. (\ref{chiEOM2}) can
be rewritten as
\begin{equation}
\dot{u}+\frac{3}{2t}u+\frac{2\gamma}{a\chi}+\frac{\rho_{b}}{\sigma}+6\pi\sigma=0.\label{u-1}
\end{equation}
By eq. (\ref{chi-1}), the physical radius of the wall is $a\chi\approx a\left(\chi_{i}-H_{i}^{-1}\right)+2t$.
If $\chi_{i}$ is sufficiently large ($\chi_{i}\gg H_{i}^{-1}$),
the third term in eq. (\ref{u-1}) can be neglected compared to the
second one. Then $u$ has an analytic solution
\begin{equation}
u(t)\approx\left[\gamma_{i}t_{i}^{3/2}+\frac{2}{5}\left(\frac{\rho_{b}}{\sigma}+6\pi\sigma\right)\left(t_{i}^{5/2}-t^{5/2}\right)\right]t^{-3/2}.\label{y-1}
\end{equation}
Let $t_{s}$ be the time when the bubble wall comes to a stop with
respect to the Hubble flow. We have $\gamma(t_{s})=1$, or $u(t_{s})=0$.
Then eq. (\ref{y-1}) gives
\begin{equation}
a_{s}\equiv a(t_{s})=\left[\frac{5\gamma_{i}}{2t_{i}}\left(\frac{\rho_{b}}{\sigma}+6\pi\sigma\right)^{-1}+1\right]^{1/5}.\label{a_s}
\end{equation}
Note that to obtain this result we have assumed $\chi_{i}\gg H_{i}^{-1}$.
Due to the third term in eq. (\ref{u-1}), which tends to slow down
the bubble, the actual $a_{s}$ could be significantly smaller. For
more general cases we need to numerically solve eq. (\ref{chiEOM2})
or eq. (\ref{u-1}) in order to find out the bubble wall's trajectory
before $t_{s}$. 

After $t_{s}$, the bubble turns around for an FRW observer, and runs
into the interior fluid. In some situations (see main text) the interior
can be approximately described by an FRW universe dominated by radiation
($\rho_{r}$) and a vacuum ($\rho_{b}$). In this case we can
still use junction conditions to match the interior and exterior regions,
then eqs. (\ref{eq:a1prime}), (\ref{eq:a2prime}) and (\ref{chiEOM2})
become
\begin{equation}
a_{1}^{\prime}=\frac{4\pi\sigma\sqrt{1-a^{2}\dot{\chi}^{2}}}{\dot{\chi}^{2}},\label{eq:a1prime-1}
\end{equation}
\begin{equation}
a_{2}^{\prime}=-\frac{4\pi\sigma a^{2}}{\sqrt{1-a^{2}\dot{\chi}^{2}}},\label{eq:a2prime-1}
\end{equation}
\begin{equation}
\ddot{\chi}+\left(4-3a^{2}\dot{\chi}^{2}\right)H\dot{\chi}+\frac{2}{a^{2}\chi}\left(1-a^{2}\dot{\chi}^{2}\right)=-\left(\frac{\rho_{b}}{\sigma}-6\pi\sigma\right)\frac{\left(1-a^{2}\dot{\chi}^{2}\right)^{3/2}}{a}.\label{chiEOM2-1}
\end{equation}
Here $a$ becomes the scale factor inside the bubble, and $a_1,a_2$ become the metric functions outside. $a$ and $H$ are now
determined by the Friedmann equations,
\begin{equation}
H^{2}=\frac{8\pi}{3}\left(\rho_{b}+\rho_{r}^{(in)}\right),
\end{equation}
\begin{equation}
\frac{\ddot{a}}{a}=\frac{8\pi}{3}\left(\rho_{b}-\rho_{r}^{(in)}\right),
\end{equation}
where $\rho_{r}^{(in)}(t)$ is the interior radiation density near
the bubble wall. 

Furthermore, we can use eq. (\ref{eq:a1prime-1}) and (\ref{eq:a2prime-1})
to find the mass measured right outside the bubble. Here we use the
standard Misner-Sharp quasi-local mass to characterize the total mass
enclose by a sphere \cite{Misner:1964je, Hayward:1994bu}. In our coordinates, the Misner-Sharp mass
right outside the wall is 
\begin{eqnarray}
M & = & \frac{a_{2}\chi}{2}\left(1-\frac{\left(a_{2}\chi\right)^{\prime2}}{a_{1}^{2}}+\dot{\left(a_{2}\chi\right)^{2}}\right)\\
 & = & \frac{1}{2}H^{2}\left(a\chi\right)^{3}+\frac{4\pi\sigma\left(a\chi\right)^{2}}{\sqrt{1-\left(a\dot{\chi}\right)^{2}}}+\frac{4\pi\sigma H\dot{\chi}}{\chi\sqrt{1-\left(a\dot{\chi}\right)^{2}}}\left(a\chi\right)^{4}-8\pi^{2}\sigma^{2}\left(a\chi\right)^{3}\label{mass}\\
 & = & \frac{4}{3}\pi\left(\rho_{r}+\rho_{r}\right)R^{3}+4\pi\sigma\gamma R^{2}+4\pi\sigma H\sqrt{\gamma^{2}-1}R^{3}-8\pi^{2}\sigma^{2}R^{3},
\end{eqnarray}
where $R=a\chi$ is the physical radius of the bubble and $\gamma$
is the bubble wall's Lorentz factor for an interior FRW observer.
Here the four terms are the volume energy, surface energy, surface-volume
binding energy and surface-surface binding energy, respectively. The
black hole mass can be determined by reading the value of $M$ when
the bubble's radius decreases to the Schwarzschild radius, i.e., $R=2M$.
The black hole mass obtained in this way is more precise than eq.
(\ref{eq:limit2_M_sub}), but we have verified that the difference
can be neglected when computing the mass spectrum of the black holes.

\section*{Acknowledgments}
I would like to thank Alex Vilenkin for stimulating discussion and comments. This work is supported by the U.S. Department of Energy, Office of High Energy Physics, under Award No. DE-SC0019470 at Arizona State University.

\bibliography{reference}
\end{document}